# Derivative relations between electrical and thermoelectric quantum transport coefficients in graphene

Xinfei Liu[1], Zhongshui Ma[2], and Jing Shi[1]

1. Department of Physics and Astronomy, University of California, Riverside, CA 92521, USA

2. School of Physics, Peking University, Beijing 100871, P.R. China

We find that the empirical relation between the longitudinal and Hall resistivities (i.e. $R_{xx}$ and $R_{xy}$) and its counterpart between the Seebeck and Nernst coefficients (i.e. $S_{xx}$ and $S_{xy}$), both originally discovered in two-dimensional electron gases, hold remarkably well for graphene in the quantum transport regime except near the Dirac point. The validity of the relations is cross-examined by independently varying the magnetic field and the carrier density in graphene. We demonstrate that the pre-factor, $\alpha_s$, does not depend on carrier density in graphene. By tuning the carrier mobility therefore the degree of disorders, we find that the pre-factor stays unchanged. Near the Dirac point, different mechanisms at low densities such as carrier localization may be responsible for the breakdown of these relations.

The empirical relation between the longitudinal resistivity $R_{xx}$ and the Hall resistivity $R_{xy}$, i.e.

$$R_{xx} = \alpha_r \cdot \frac{B}{n} \frac{dR_{xy}}{dB}, \qquad (1)$$

where $n$ is the carrier density, $B$ the magnetic field and $\alpha_r$ a $B$-independent but sample dependent constant, was first discovered by Chang and Tsui[1] in a two-dimensional electron gas (2DEG), and later on found to hold remarkably well in a number of other 2DEG systems in both integer and fractional Hall regimes.[2,3,4,5] Although the microscopic origin of this empirical relation is not completely understood, it has been proposed[6,7] that due to inhomogeneities existing on different length scales, local fluctuations in the Hall resistivity via carrier density fluctuations are linked to the global longitudinal resistivity. More recently, Tieke et al.[8] experimentally demonstrated another important relation between the longitudinal ($S_{xx}$) and transverse ($S_{yx}$) thermoelectric transport coefficients, i.e.

$$S_{yx} = \alpha_s \cdot \frac{B}{n} \frac{dS_{xx}}{dB}, \qquad (2)$$

in a 2DEG under a varying magnetic field. In the meantime, they showed that this relation is equivalent to that of the resistivity under certain conditions. This raises further interesting questions about the common physical origin of both relations. For example, both $S_{xx}$ and $S_{yx}$ are usually dominated by the phonon drag contribution at low temperatures in 2DEG systems.[9] The fact that these relations hold for thermoelectric transport coefficients in 2DEGs suggests that the validity of relations seems to be broader.

In conventional 2DEG systems, the empirical relations were originally found and subsequently verified for fixed $n$ and a varying $B$-field. In principle, both carrier density and carrier type can be tuned by electrostatic gating in some devices. In the quantum Hall regime, since the filling factor is $\nu = \frac{nh}{eB}$, the two relations can also be expressed by

$$R_{xx} = -\alpha_r \frac{e}{C_g} \frac{dR_{xy}}{dV_g}, \qquad (3)$$

$$\text{and } S_{yx} = -\alpha_s \frac{e}{C_g} \frac{dS_{xx}}{dV_g}, \quad \text{or} \quad S_{xy} = \alpha_s \frac{e}{C_g} \frac{dS_{xx}}{dV_g}, \qquad (4)$$

where $e$ is the absolute value of the electron charge, $C_g$ is the gate capacitance, $\alpha_s$ is an $n$-independent constant, and $V_g$ is the gate voltage that controls the carrier density and carrier type. Hence, it is expected that these relations hold when $n$ is tuned instead.[10] By varying $n$, the relative length scale of local inhomogeneities can be changed and therefore the effect of local density fluctuations on global dissipation can be investigated, which offers a different perspective to examine the validity of the derivative relations. Because of its gapless excitation spectrum, graphene is such an ideal 2D electron system for this purpose.[11,12] First of all, the charge carriers can be smoothly tuned from electrons to holes through the Diract point. Second, graphene has a distinct Dirac dispersion relation, which could bear unusual consequences in both electrical and thermoelectric transport properties as well as the derivative relations. Furthermore, the charged impurity state and consequently the carrier mobility can be tuned in the same devices so that the relations can be examined as the mobility is varied while leaving everything else untouched.[13] Existing graphene thermoelectric transport results seem to suggest that the diffusion thermopower dominates,[14,15,16] which is different from the situation in conventional 2DEGs.[9] In quantizing magnetic fields, we do not observe any dramatic temperature dependent conductivity or thermopower predicted for phonon drag for conventional 2DEGs.[17] Here we report our experimental study on both electrical and thermoelectric transport coefficients and their derivative relations in the quantum Hall regime as the carrier density and magnetic field are varied independently. We further investigate the validity of the derivative relations in the same graphene devices with a variable mobility tuned by using a unique method we discovered earlier.

Single-layer graphene devices are fabricated on $SiO_2$/Si with electron beam lithography as reported previously.[15,18] Multiple contacts and a micro-heater close to the graphene flake (inset of Fig. 2a) allow us to measure both electrical and thermoelectric coefficients in longitudinal and transverse directions. For thermoelectric measurements, between heater-on and -off, a typical temperature difference $\Delta T$ between two thermo-emf leads that are ~ 10 μm apart is ~ 200 mK, which is calculated from the four-terminal resistance change of the resistive thermometers (i.e. contacts 1 & 6). The same thermometer leads are also used to measure the change in thermo-emf, $\Delta V_x$, or the corresponding electric field $E_x$. Then the longitudinal thermoelectric or Seebeck coefficient $S_{xx}$ is determined by $S_{xx} = \frac{E_x}{(\nabla T)_x} = -\frac{\Delta V_x}{\Delta T}$. The measured temperature gradient, $(\nabla T)_x$, along with the transverse thermo-emf $\Delta V_y$ from contacts 2 & 4 or 3 & 5, is used to calculate the transverse thermoelectric or Nernst coefficient $S_{yx}$, by $S_{yx} = \frac{E_y}{(\nabla T)_x} = -\frac{L}{W} \cdot \frac{\Delta V_y}{\Delta T} = -S_{xy}$,[19] here $L$ and $W$ being the spatial separations between the longitudinal and transverse emf leads, respectively. The measurements were carried out in an Oxford cryostat which covers the temperature range from 1.5 to 300 K and magnetic fields up to 8 T. In this letter, we report the data from two representative single-layer devices (devices 1 & 2) whose mobility is tuned by over a factor of 3 using the method described previously.[13] We only include low-temperature data which show clear quantum Hall plateaus and quantum oscillations.

Fig. 1 shows 2D plots of $R_{xx}$, $R_{xy}$, $S_{xx}$, and $S_{xy}$ vs. $V_g$ and $B$ at 20 K for device 1. Each pair, $R_{xx}$ and $R_{xy}$, or $S_{xx}$ and $S_{xy}$, is taken simultaneously under exactly the same conditions as $V_g$ is swept at fixed $B$-fields. A number of $V_g$-dependent curves for different $B$-fields are used to generate these plots. The Dirac point in the raw data is located at ~ +2 V, but in the figures $V_g$ plotted relative to the Dirac point for clarity. Multiple plateaus or quantum oscillations corresponding to different Landau levels (LL) can be clearly identified from these plots. At $R_{xx}$ peaks where the chemical potential passes LLs, $S_{xx}$ shows either peaks or dips, with peaks corresponding to hole LLs and dips corresponding to electron LLs. In quantum Nernst signal $S_{xy}$, there is a main positive peak right at the Dirac point or the zeroth LL, which is accompanied by a pair of side peak-dip feature at each LL on both sides.[14,15,16]

To see how various transport coefficients are related, we plot measured $R_{xx}$ and $S_{xy}$ vs. $B$ for a fixed $V_g$. Data shown in Fig. 2 are for $V_g$= +25 V and $T$= 20 K. We choose the data at this temperature because the thermoelectric coefficients have a larger signal-to-noise ratio. Under these conditions, all transport coefficients show characteristic features such as the half-integer quantum Hall effect in $R_{xy}$ (not shown here) and quantum oscillations in other coefficients. At this temperature, $R_{xx}$ clearly does not vanish at the half-integer fillings. Both experimentally measured $R_{xx}$ ($S_{xy}$) and calculated $R_{xx}^{Calc}$ ($S_{xy}^{Calc}$) are displayed in Fig. 2a (2b), where $R_{xx}^{Calc}$ and $S_{xy}^{Calc}$ are obtained from measured $R_{xy}$ and $S_{xx}$ by applying Eqs. 1 and 2, respectively. Clearly, the main oscillatory features in both $R_{xx}$ and $S_{xy}$ correspond well to those in $R_{xx}^{Calc}$ and $S_{xy}^{Calc}$ calculated from independently measured $R_{xy}$ and $S_{xx}$. The excellent correspondence justifies the validity of derivative relations for both electrical and thermoelectric transport coefficients at this representative electron density ($\sim 1.7 \times 10^{12}$ cm$^{-2}$) for graphene whose charge carriers are massless Dirac fermions.

The continuously gate tunable carrier density and the bipolar transport are a unique property of graphene. By sweeping $V_g$, not only different LLs, but also the Dirac point, can be easily accessed, which offers a convenient way to examine the validity of the relations. The 4 T line scans in Fig. 1 are selected to be re-plotted in Fig. 3. Figs. 3a and 3b show the same two pairs of transport coefficients as in Figs. 2a and 2b, except that they are now plotted as functions of $V_g$ for $B=$ 4 T. To better compare the oscillatory features that have smaller magnitude, we exclude the central peak regions and display them separately in the insets. At 4 T, up to six LLs on each side of the Dirac point can be readily identified and labeled, indicating very high sample quality (the mobility of the device in this state is ~12900cm$^2$/Vs). Note here that the peaks in $R_{xx}$ in Fig. 3a correspond to the zeros in quantum Nernst signal in Fig. 3b. We apply Eqs. 3 and 4 to calculate $R_{xx}^{Calc}$ and $S_{xy}^{Calc}$ from the independently measured Hall and Seebeck coefficients, and then show them in the same figures. Except for the zero-th LL, features in $R_{xx}$ and $S_{xy}$ for all other LLs match very well with the respective features in $R_{xx}^{Calc}$ and $S_{xy}^{Calc}$. From the single particle picture, we expect $R_{xy}$ to switch the sign from negative to positive as $V_g$ sweeps through the Dirac point from the negative side (i.e. from holes to electrons), which gives rise to negative $R_{xx}^{Calc}$ at the Dirac point according to Eq. 3. This disagrees with actually measured $R_{xx}$ and is clearly unphysical. Similarly, as $V_g$ is swept through the Dirac point from the negative side, $S_{xx}$ passes zero from positive to negative and this consequently produces negative $S_{xy}^{Calc}$ at the Dirac point according to Eq. 4. It also contradicts with the measured $S_{xy}$ that has a positive peak. Interestingly, both Eqs. 3 and 4 fail at the zero-th LL. We will discuss more about this later. Apart from the zero-th LL, the agreement is good in both polarity and the decreasing trend of the oscillation amplitude at higher densities. We also note that while the match is perfect on the electron side, there is a phase shift between the measured and calculated curves on the hole side. We also observe similar phase shift at different magnetic fields.

In the original work of Chang & Tsui,[1] the equation contains $n$ in the denominator, which is in the same form as our Eq. 1. In the same spirit, Eq. 2 also contains $n$. Whether this $n$ appears in Eqs. 1 and 2 does not have any consequence if $n$ is fixed while $B$ is swept in experiments. In later papers,[2,3,4,5] this $n$ was dropped from the equations. In our experiments, however, we sweep $V_g$ so that $n$ is a variable. Obviously, the presence or absence of $n$ does have important consequences. Eqs. 3 and 4 are equivalent to Eqs. 1 and 2; therefore, $n$ cancels out. In Figs. 3a and 3b, both $R_{xx}^{Calc}$ and $S_{xy}^{Calc}$ are calculated with Eqs. 3 and 4, i.e. without $n$ in front. To distinguish the two cases, we have also calculated $R_{xx}^{Calc}$ and $S_{xy}^{Calc}$ with $n$ in Eqs. 3 and 4 (not shown in the figure). We find that by including $n$ in front, the oscillation amplitude of both $R_{xx}^{Calc}$ and $S_{xy}^{Calc}$ stays approximately constant, which obviously disagree with the observed decreasing trend in the amplitude of $R_{xx}$ and $S_{xy}$ as $n$ increases (in Figs. 3a and 3b). Hence we conclude that Eqs. 1-4 are consistent with our experimental data; therefore, both pre-factors $\alpha_r$ and $\alpha_s$ defined in our way are independent of $n$. We have used the $n$-independent pre-factors extracted from our data to calculate the dimensionless pre-factors in Tieke's work for the particular carrier density of their samples,[8] and found reasonable agreement with each other.

We have just demonstrated that the empirical relations hold very well for high LLs. At low densities, e.g. approaching the zero-th LL, the relations cease working as discussed earlier. To focus on low densities, we sweep the magnetic field at a fixed $V_g$ instead. Fig. 4a shows the comparison of $R_{xx}$ and $R_{xx}^{Calc}$ for $V_g$= +2 V (at the Dirac point) in a varying magnetic field. $R_{xx}$ monotonically increases as $B$ increases, but in the meantime $R_{xy}$ only increases slightly so that $R_{xx}^{Calc}$ stays nearly constant. Apparently, near the Dirac point, the behaviors of $R_{xx}$ and $R_{xy}$ are qualitatively different from those at higher densities, and the derivative relation does not hold. The increasing trend in $R_{xx}$ observed here is similar to the high-field insulating behavior reported previously.[20] Therefore, the relations are likely complicated or even rendered invalid by possible localization occurring at the Dirac point. Further investigation of the zero-th LL behavior is obviously beyond the scope of this work.

We are able to tune the low-temperature mobility by controlling and freezing different high-temperature charge environments of the same graphene device.[13] We have performed the mobility tuning in devices 1 and 2. Here we show the results from device 2, in which we have succeeded in setting four different mobility values and obtaining a complete set quantum Seebeck and Nernst data at low temperatures. In general, the higher the mobility, the better quality data the device has. We find that the derivative relation between $S_{xx}$ and $S_{xy}$ (i.e. Eq. 4) holds well for all mobility states away from the Dirac point. From those results, we obtain the pre-factor, $\alpha_s$, for four different mobility values, as shown in Fig. 4b. Apparently, parameter $\alpha_s$ does not change significantly with the carrier mobility. In graphene, one school of thoughts is that the mobility is largely determined by the charged impurities that have a direct consequence on the carrier density fluctuations near the Dirac point.[21,22,23,24] By varying carrier mobility, the fluctuations in carrier density and their length scales may be varied; therefore, one might expect parameter $\alpha_s$ to be different. Our results indicate that within the tunable range of the mobility, neither the derivative relations nor the pre-factor is obviously affected. The relationship between the pre-factor and the carrier mobility was previously studied in conventional 2DEG systems.[9] In several samples with different mobilities, it was found that the pre-factor depends on the carrier mobility in the classical transport regime at high temperatures, but becomes a mobility independent constant in the quantum transport regime at low temperatures. Our results in the same graphene device with different mobilities agree with those in conventional 2DEG in the quantum transport regime.

In summary, we have measured four electrical and thermoelectric transport coefficients in quantizing magnetic fields and examined the empirical relations among them by independently varying the magnetic field and carrier density. These relations hold very well at high LLs but fail at the zero-th LL. The universality of the derivative relations suggests some common physical origin shared by all 2D electron systems in quantum transport regime despite their widely different properties.

This work was supported by DMEA/CNN H94003-10-2-1004 (XFL and JS), DOE DE-FG02-07ER46351 (JS) and NNSFC 91021017 (ZSM).

Figure captions:

Fig. 1. 2D plots of four measured transport coefficients of a graphene device with a varying gate voltage and a varying magnetic field: $R_{xx}$ (a), $R_{xy}$ (b), $S_{xx}$ (c), and $S_{xy}$ (d). Each pair, $R_{xx}$ and $R_{xy}$, or $S_{xx}$ and $S_{xy}$, were measured simultaneously, and all four coefficients were measured at T= 20 K. The gate voltage is referenced to the Dirac point which is located at ~ +2 V.

Fig. 2. (a) Experimentally measured $R_{xx}$ and calculated $R_{xx}^{Calc}$ (using Eq. 1) from $R_{xy}$ vs. magnetic field. (b) Experimentally measured $S_{xy}$ and calculated $S_{xy}^{Calc}$ (using Eq. 2) from $S_{xx}$ vs. magnetic field. The inset shows a scanning electron microscope image of a graphene device.

Fig. 3. (a) Experimentally measured $R_{xx}$ and calculated $R_{xx}^{Calc}$ (using Eq. 3) from $R_{xy}$ vs. gate voltage at B = 4 T and T= 20 K. The peaks are labeled by LL indices. (b) Experimentally measured $S_{xy}$ and calculated $S_{xy}^{Calc}$ (using Eq. 4) from $S_{xx}$ vs. gate voltage under the same conditions as in (a). The insets in both (a) and (b) are the data near the Dirac point.

Fig. 4. (a) Comparison of $R_{xx}$ and $R_{xx}^{Calc}$ vs. B at $V_g$= +2 V, at the Dirac point. (b) $\alpha_s$, the pre-factor in Eq. 4, as a function of carrier mobility $\mu$ of a device.

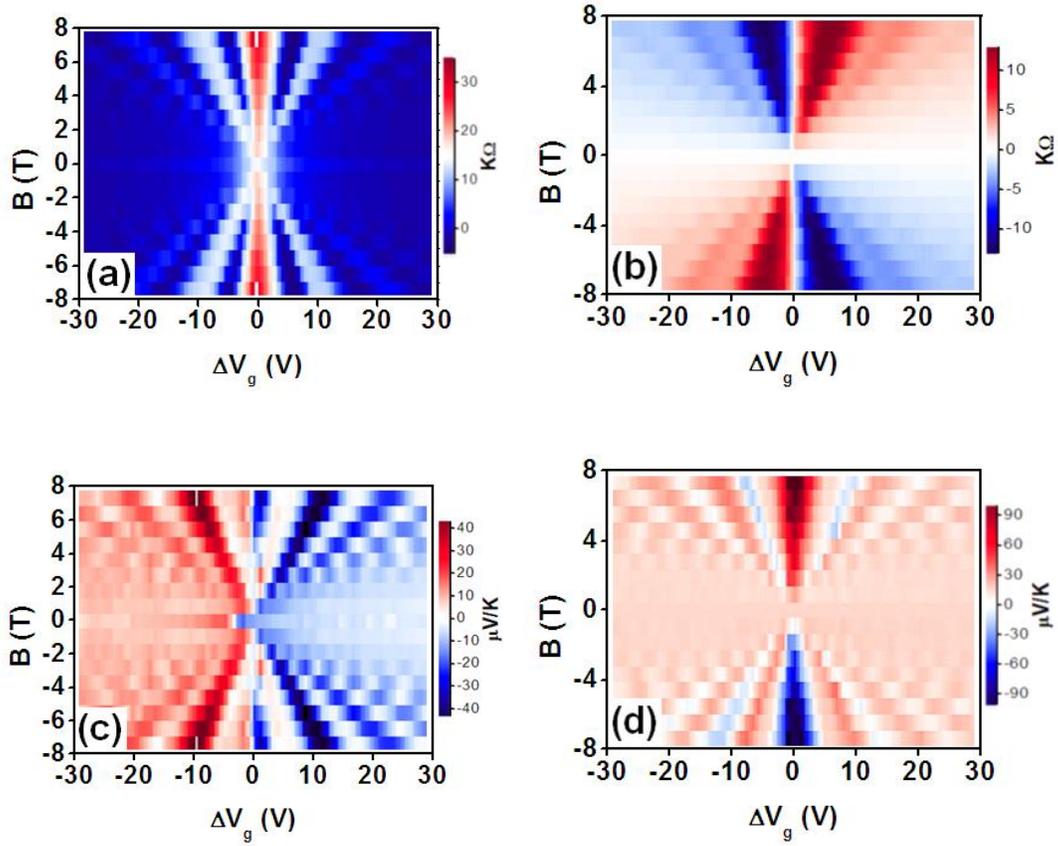

Figure 1

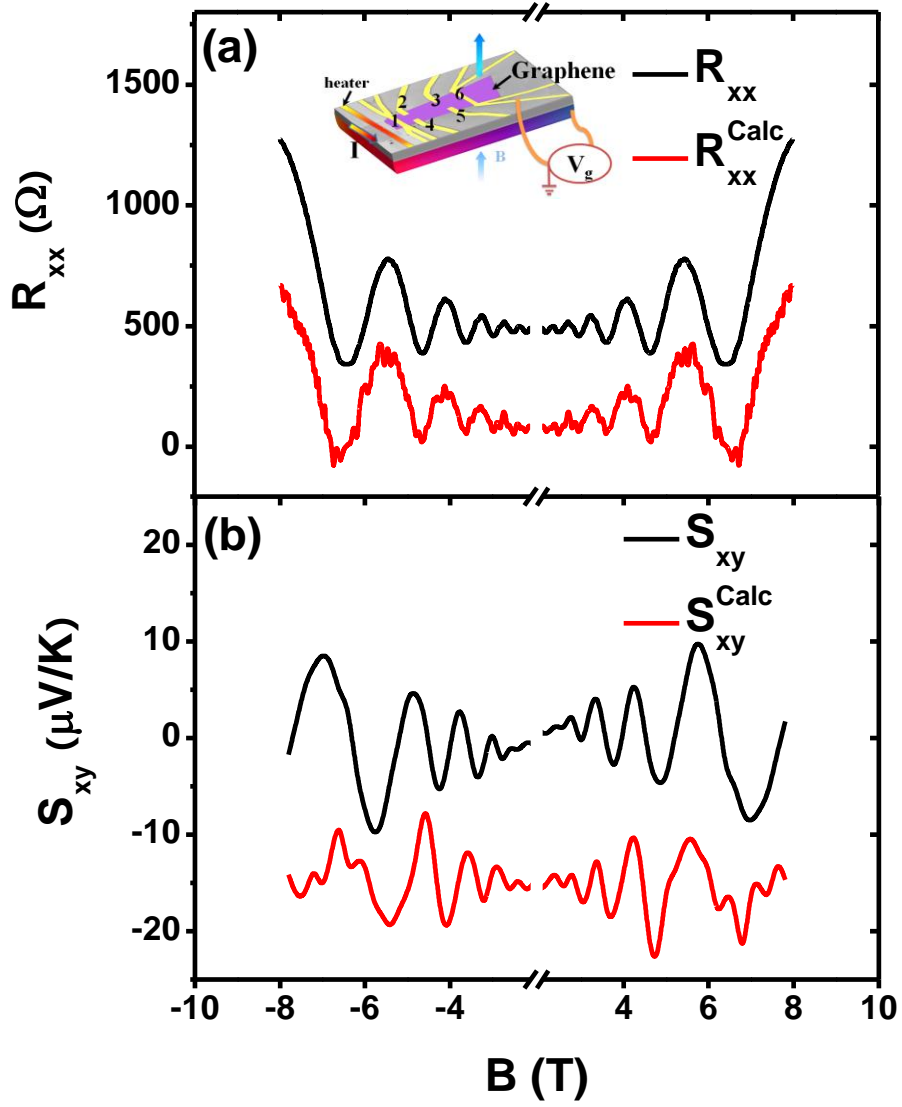

Figure 2

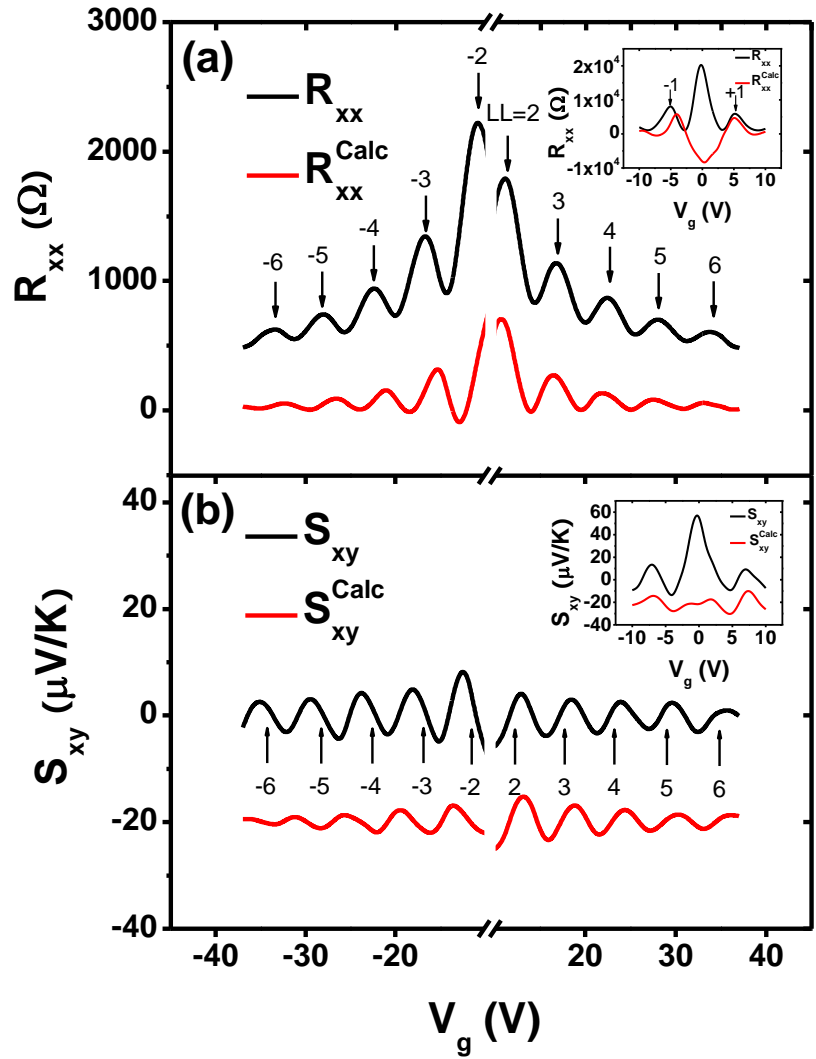

Figure 3

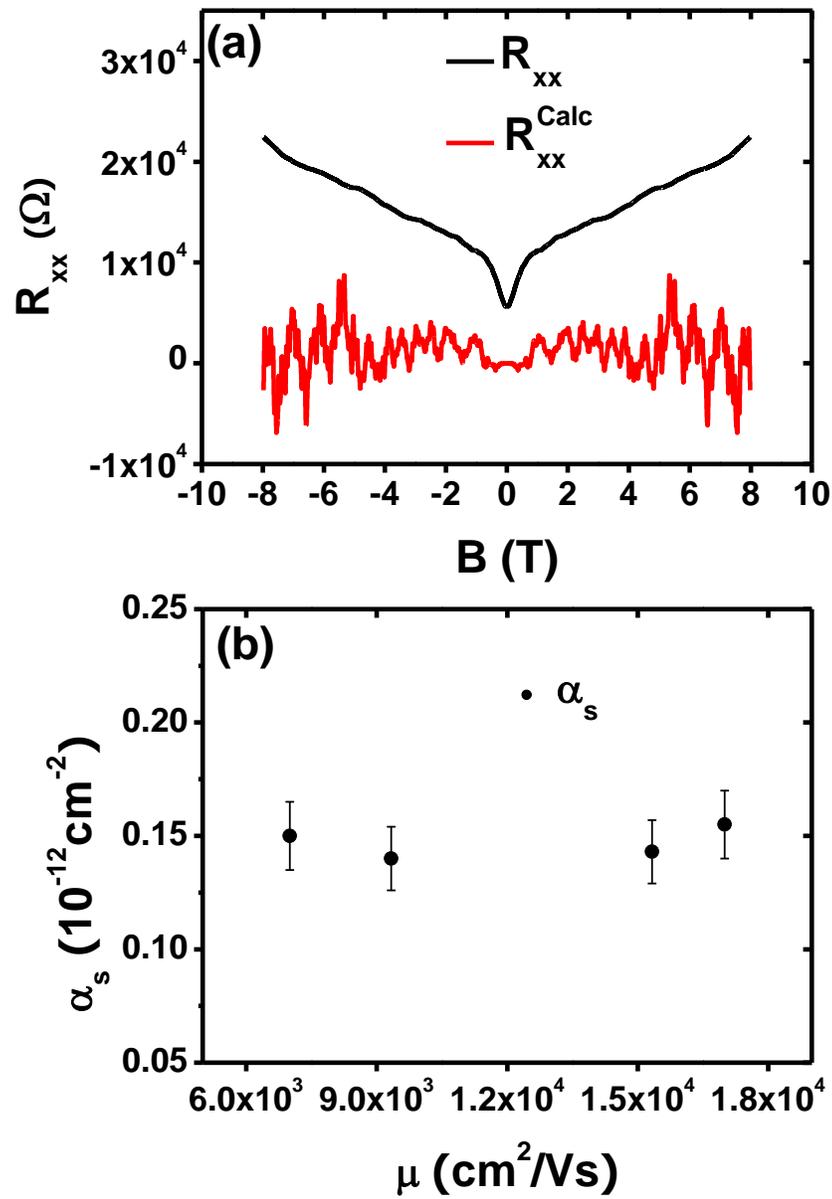

Figure 4